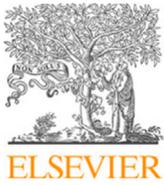
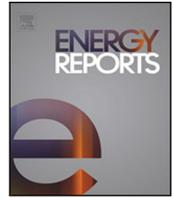

Research paper

# An artificial neural network based approach for harmonic component prediction in a distribution line

Dixant Bikal Sapkota *, Puskar Neupane, Kajal Pokharel, Shahabuddin Khan

*Department of Electrical Engineering, Pulchowk Campus, IOE Tribhuvan University, Pulchowk, Lalitpur, 44600, Nepal*

ARTICLE INFO

ABSTRACT

With the increasing use of nonlinear devices in both generation and consumption of power, it is essential that we develop accurate and quick control for active filters to suppress harmonics. Time delays between input and output are catastrophic for such filters which rely on real-time operation. Artificial Neural Networks (ANNs) are capable of modeling complex nonlinear systems through adjustments in their learned parameters. Once properly trained, they can produce highly accurate predictions at an instantaneous time frame. Leveraging these qualities, various complex control systems may be replaced or aided by neural networks to provide quick and precise responses. This paper proposes an ANN-based approach for the prediction of individual harmonic components using minimal inputs. By extracting and analyzing the nature of harmonic component magnitudes obtained from the survey of a particular area through real-time measurements, a sequential pattern in their occurrence is observed. Various neural network architectures are trained using the collected data and their performances are evaluated. The best-performing model, whose losses are minimal, is then used to observe the harmonic cancellation for multiple unseen cases through a simplified simulation in hardware-in-the-loop. These neural network structures, which produce instantaneous and accurate outputs, are effective in harmonic filtering.

## 1. Introduction

Power quality in an AC system is the degree to which voltage and current waveforms, of the power frequency, resemble a sinusoidal shape. It is crucial to maintain high levels of power quality to ensure the proper functioning of connected devices, reduction of unwanted losses and maintenance of system reliability. Harmonics, in a power system, refers to the sinusoidal components of voltage and current waveforms that have a higher frequency, which are integer multiples of the power frequency. In a power system designed to operate at 50 Hz, 3rd harmonics represent components oscillating at 150 Hz, 5th harmonics represent components operating at 250 Hz, and so on. The degradation of waveform shapes caused by the addition of such high-frequency components can be seen in Fig. 1. Harmonics primarily result from the operation of non-linear loads and are highly undesirable as they lower power quality causing damage to various equipment. Total Harmonic Distortion (THD) is a unit of harmonic measurement and is given by Eqs. (1) and (2) for current and voltage waveforms respectively.

$$THD_i(\%) = \frac{\sqrt{I_2^2 + I_3^2 + \cdots + I_n^2}}{I_1} * 100\% \quad (1)$$

$$THD_v(\%) = \frac{\sqrt{V_2^2 + V_3^2 + \cdots + V_n^2}}{V_1} * 100\% \quad (2)$$

A higher value of THD signifies a larger presence of harmonic magnitudes, i.e., waveforms of non-fundamental frequencies. Various standards and regulations define the limits of harmonic distortions for any system to ensure safe and reliable operation. IEEE 519 (2022) is one of the standards that establishes the goals for the design of electrical systems comprising linear and nonlinear (harmonic-producing) loads. Growth in the use of power electronic devices, such as diodes, transistors, converters, rectifiers and others, that primarily behave as nonlinear loads, has raised significant concerns regarding power quality. Though the individual devices have diminutive effects on any large power grid, thousands being operated simultaneously will certainly have adverse effects. Further, modern industrial machinery that utilizes Variable Frequency Drives (VFDs), which are used for accurate speed control of electric motors, and integration of Variable Renewable Energy Resources (VERs) like solar and wind, have significantly contributed to the distortion of overall grid parameters.

Over the years, various methodologies have been adopted to suppress and minimize harmonic components. Quality assurance checks,

* Corresponding author.

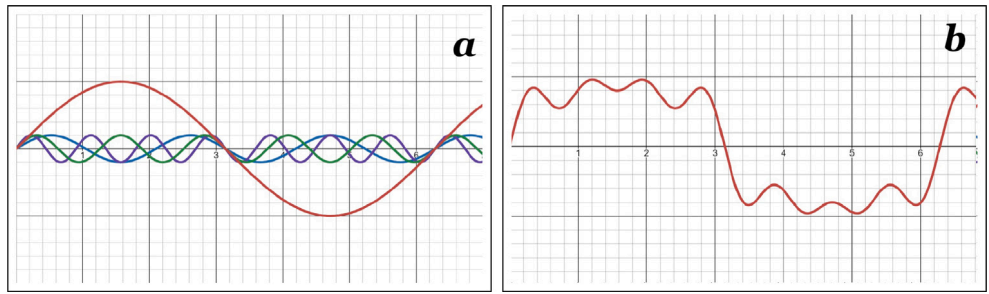

**Fig. 1.** (a) Fundamental waveform, in red, and other higher frequency harmonics (3rd, 5th and 7th), (b) Resulting degradation in waveform due to addition of these harmonics.

standardization of equipment, and usage of passive and active filters have made significant improvements in the overall power quality. In addition, other techniques like placement and grouping of nonlinear loads, use of transformers with special connections and appropriate earthing techniques have also aided in the suppression of harmonics. Due to the changing face of modern power systems, with a much higher number of nonlinear devices both on the load side and source side, active filters are highly preferable. The main advantage of active filters over passive ones is their fine response to changing loads and harmonic variations (Fuchs and Masoum, 2023). This is done through real-time measurements and complex transformations by the control mechanism of the filter. These actions take time because resulting in previous measurements influencing current outputs, prompting an error in the filtering procedure. The distortion term being out of date resulting in a distortion cancellation error is the biggest drawback of any active filter. This calls for a system capable of instantaneous yet accurate response, which can be provided by an Artificial Neural Network (ANN) based approach.

ANNs are capable of modeling a wide range of complex nonlinear systems because of their inherent features (Suykens et al., 2012). A single node, called perceptron (Rosenblatt, 1958), in a larger network can be seen in Fig. 2. The particular node here takes three inputs ($x_0$, $x_1$, and $x_2$), and gives a single output ($y$). The inputs are multiplied by their corresponding weights ($w_0$, $w_1$, and $w_2$) then added together. The sum value then gets passed into the activation unit which has a nonlinear function ($act(x)$) embedded inside. These non-linear functions may be sigmoidal (Eq. (3)), tanh (Eq. (4)) or ReLU (Eq. (5)), which are some of the common ones. The output of this second unit, $act(\sum(x \cdot w))$, is the output of the node. During the training process, the node continually adjusts its weights ($w$) to improve its performance. The nonlinear nature of the embedded activation function inside each node allows it to accurately model a nonlinear system. In a complex neural network, numerous such nodes are present which are interconnected to each other. Special nodes, such as in a Recurrent Neural Network, can capture sequential patterns by holding information from previous inputs. Modern networks, like Transformers, can contextualize the meaning of a sequence through a mechanism called *Attention*. Regardless of the nature of a model, with proper training on a relevant dataset, they can produce accurate outputs, which may be in the form of probability, regression or classification. While the training process may be time-consuming, which largely depends on training data and model architecture, ANNs produce output instantaneously which can be leveraged in a real-time control system such as an active filter.

$$Sigmoidal, \ \sigma(x) = \frac{1}{1 + e^{-x}} \tag{3}$$

$$tanh(x) = \frac{e^x - e^{-x}}{e^x + e^{-x}} \tag{4}$$

$$Relu(x) = max(0, x) \tag{5}$$

Taking advantage of such properties of neural networks, our work presents a method to predict individual harmonic components directly.

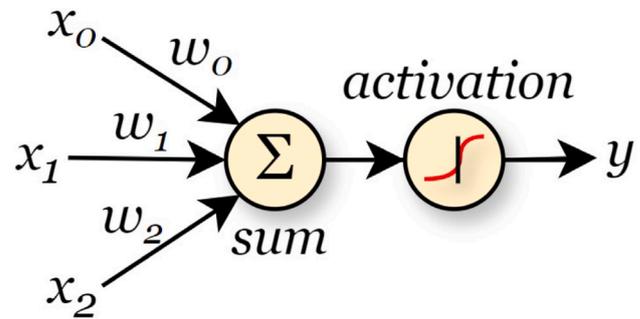

**Fig. 2.** A single neural network node which takes three inputs (x) and gives a single output (y).

This significantly reduces the complexity of control algorithms required and fundamentally removes any concerns of time delay. On top of this, exploratory data analysis can help design a sturdy architecture, requiring only the basic measurements for input such as time of day and measured current magnitude or just a self-sequential window. Additional inputs such as the number of households and total connected load can be used to address the trend of growth in demand to keep the model accurate for multiple years of use.

In this work, we explore different neural network architectures for individual harmonic component prediction to evaluate their performance using only basic inputs. While the underlying concept of every neural network may be similar, different architectures are effective at tackling different nature of problems. After ample data analysis, we implement ensemble models, feed-forward networks and RNN-based sequential models. Among them, the best-performing ones are highlighted and used to visualize waveform filtering through Hardware-in-the-Loop (HIL) simulation. The remainder of this paper contains the following sections: a review of previous and current topologies for harmonic filters, our methodology — from data collection to model building, followed by the results and conclusion.

## 2. Literature review

Power quality analysis and improvement has been a significant area of research and technological development due to an increasing usage of nonlinear devices in the form of power electronics, both in generation and consumption that need inverters and converters. Both electric utilities and end users are becoming increasingly concerned about the quality of electric power (Dugan et al., 2002). The main reasons for increased concern include newer-generation load equipment consisting of power electronics, emphasis on the use of adjustable-speed motor drives in industries, and the cascading capability of an interconnected network. The use of electrochemical rectifiers and an increase in the application of AC and DC adjustable speed drives has dictated the harmonic "pollution" in power systems (Rice, 1986). A preliminary survey conducted by Nejdawi et al. (1999) concludes the ramping of harmonic pollution in its past decade, with a mean $THD_v$ increase of 0.1% per year. With open electricity markets, a general power quality must be guaranteed by the system operator to maintain a level of security. For this, there is an urgent need to adapt existing systems to control harmonics and monitor their implementation more rigorously (Arrillaga and Watson, 2003).

Various options for improving the utility interface of power electronic equipment are enlisted by Mohan et al. (2003), which include the use of passive circuits, active shaping of waveforms, use of separate transformer-isolated filtering circuits, and different rectifier topologies. The author of the paper (Afonso et al., 2021) presents a highly detailed review of various power electronic topologies applied to address power quality issues, including filtering methods. The work enlists various developed topologies for active power filters, both shunt (Sasaki and Machida, 1971; Verdelho and Marques, 1998; Jiao et al., 2019; Kanjiya et al., 2014; Jou et al., 1994; Costa-Castelló et al., 2009; Fabricio et al., 2018) and series (Javadi et al., 2016; Chaudhari et al., 2012; Kim et al., 2004; George and Agarwal, 2006; Dixon et al., 1997; Pinto et al., 2011). Although various filters with different topologies are available for harmonic suppression, each comes with certain shortcomings. Commonly used filters often lack real-time monitoring, measurement, and prediction of harmonics, which are crucial for effective performance (Wang, 2024). Moreover, these filters are typically designed to address harmonics from specific sources, limiting their versatility. A filter capable of suppressing harmonics from diverse sources would not only improve efficiency but also offer significant economic benefits by reducing the need for multiple specialized filters.

Artificial Neural Networks (ANNs) have seen significant advances and have increasingly influenced control systems across various industries. The explosion of data availability due to digital devices, advanced sensors and cloud storage has been the driving factor for this. Other factors including technological advancements in the field of processing capabilities, algorithmic understanding, and continued demand, have also paved the way for the growth of ANNs. The major advantages of ANNs include their ability to model complex non-linear systems (Bermejo et al., 1844; Sapkota et al., 2024; Olivencia Polo et al., 2015; Zhang et al., 1998), give real-time outputs and adapt with different systems. Multiple feedforward network architectures have been explored as nonlinear models, and it has been demonstrated that they all possess strong abilities to represent and control intricate nonlinear systems (Waterworth and Lees, 2000; Chen and Billings, 1992).

Control systems have also seen significant improvement with more research and application of ANN. Integrating AI into closed-loop systems will enhance the overall performance of the control system and the controller behavior, particularly for multivariable and nonlinear processes (Schöning et al., 2022). The implementation of Self-Organizing Map (SOM) based control systems has also been proven to increase system flexibility, thus expanding application opportunities (Lightowler and Nareid, 2003). Hence, ANN has been used in various fields encompassing nonlinear systems and different control systems, achieving reliable and efficient operation. The use of ANN in power system applications is continually being explored. With neural network-based control, various parameters of the power system, such as voltage and frequency, can be kept within a stable range with seemingly instantaneous responses. Pandey et al. (2023) discusses different artificial intelligence techniques used for the efficient operation, planning, and control of power systems by regulating various parameters. Furthermore, Chen et al. (2022) provides a comprehensive review of various Reinforcement Learning (RL) methods and how their applications affect decision-making and control in power systems, in three key applications: frequency regulation, voltage control, and energy management. This work also presents critical issues in such applications, mainly relating to the system's safety, robustness and scalability.

The integration of active filtering technique and neural networks was done in Shatshat et al. (2004). High speed, accuracy, efficiency and flexibility are mentioned as the main advantages of this proposed active filter system utilizing an Adaptive Linear Neural (ADALINE) network. The work exhibits an enhancement in the filter performance due to the neural network's speedy tracking of harmonics. ADALINE network is also used to control three-phase shunt active filters (Qasim et al., 2014). Authors in paper (Lin, 2007) have developed a functional neural network model for detecting harmonics which was verified for different waveforms containing up to the 11th harmonic order. It responds quickly to the input harmonics, thereby improving the performance of harmonic compensation. Since then, through the integration of ANNs with filter control, different harmonic suppression topologies have been introduced (Iqbal et al., 2021; El-Mamlouk et al., 2011; do Nascimento et al., 2011; Guzman et al., 2016). The neural network architectures and processing capability of today's computers have skyrocketed their potential and efficiency. Effective harmonic estimations utilizing Machine Learning approaches, including neural networks, have been reviewed in Taghvaie et al. (2023).

Continuing this effort to highlight the capabilities of neural networks, this paper compares various neural network architectures that can be used to accurately predict individual harmonic component magnitudes in a given distribution network. The purpose is to provide instantaneous output with minimal inputs. Additionally, these predicted magnitudes can be used to generate reference signals for the control of an active filter in a distribution network.

## 3. Methodology

This research focuses on developing an optimal neural network architecture capable of predicting the magnitudes of harmonic components with minimal and very basic inputs. To achieve this a relevant dataset is needed, along with a proper validation methods. The data used in this research is collected using a power analyzer at the survey location, which is capable of logging many grid parameters at once. The implemented neural networks, after training, validate their predictions by comparing them with the measured values. Furthermore, to observe the waveforms before and after filtering in an oscilloscope, a simplified filter is simulated in the Typhoon-HIL 404 environment. The predictions outputted by the best-performing model are utilized for this simulation. To detail the overall process, this section has been divided into three parts: *Data Collection*, *Data Analysis* and *Model Architecture Design*.

### 3.1. Data collection

A part of Pulchowk Campus, located at Lalitpur, Nepal which covers an area of 45,000 m² as seen in Fig. 3, is selected as the survey area. The area comprises two hostels, a canteen and a large staff living quarter.

A power analyzer, Elcontrol NanoVIP, is used to measure several grid parameters at the secondary side bus bar of the distribution transformer. The device is capable of measuring various information: Voltage, Current, THD, Power (Active and Reactive), Energy, Frequency and Power Factor for all three phases and neutral. All the recorded information, logged every 30 seconds, is then exported to a comma-separated values (CSV) file for further processing and analysis. The power analyzer is capable of measuring up to 50th harmonic order. However, only 3rd, 5th and 7th harmonics are utilized for further analysis as they are the most significant in all cases. The magnitudes of these harmonic components are extracted separately, analyzed and later concatenated with the previously logged data. The device calculates individual harmonics through Fourier transformation. It is represented by $F(w)$, given by Eq. (6) for any waveform in the time domain, $f(t)$.

$$F(w) = \int_{-\infty}^{\infty} f(t).e^{-jwt}dt \tag{6}$$

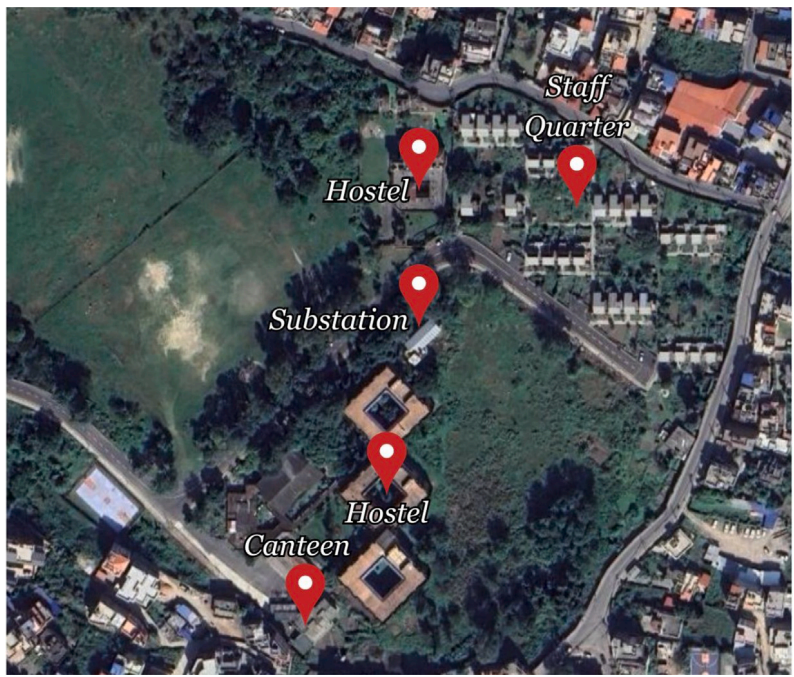

**Fig. 3.** The local campus area surrounding the substation where the power analyzer is connected.

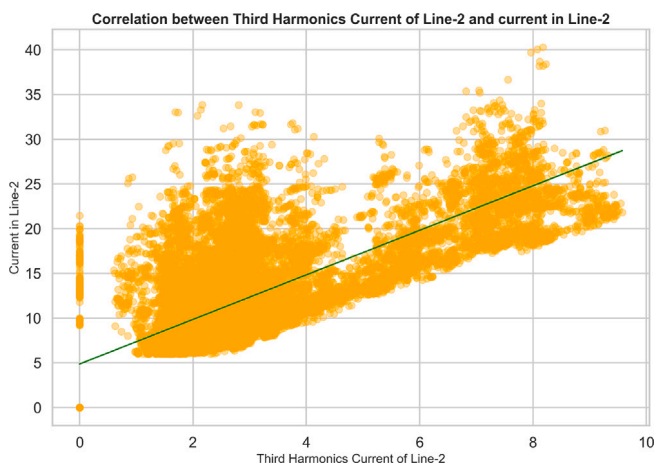

**Fig. 4.** Correlation of Third harmonics component and line current for Line-2.

The logged data is collected daily, and extracted separately. These files are exported to CSV files individually and later concatenated to form the overall dataset. Cleaning the dataset plays an important role in ensuring that the models are trained and evaluated properly. Anomalies in the collected data, particularly due to power outages, visible equipment errors, and extremely low loading are removed.

### 3.2. Data analysis

The overall dataset, in the CSV format, is then analyzed to understand any patterns or relations among the different parameters. The voltage harmonics are substantially low compared to the current harmonics in our data. So, any harmonic components mentioned by default refer to the current waveform components. Initially, a scatter plot between third harmonics and line current is plotted as seen in Fig. 4, which shows a somewhat positive general correlation.

It is important to understand if there exists any underlying relation among the harmonic components themselves. Fig. 5 shows the scatter plots, visualizing the relation between the harmonic components. The first plot shows a correlation between the 3rd and 5th components, the second plot between the 3rd and 7th components and the third is plotted between the 5th and 7th components. We can easily observe the highly linear nature of the third plot signifying a stronger correlation between 5th and 7th components.

To observe the nature of THDi throughout any given day, the data is segmented on the basis of the dates and plotted at three different time groups for a single date, as seen in Fig. 6. The mean THDi for the entire dataset is represented by a faint horizontal dotted line. It is observed that the THDi is higher during the mornings and evenings than during the afternoons. This can be accredited to an extensive use of induction cookers and other power electronic devices when most family members are present at home.

Auto-correlation measures the relationship between any variable's present and past values. A plot showing auto-correlation can help identify sequential patterns highlighting the influence of past time steps on future time steps. When plotted for the third harmonic component of line-2, as seen in Fig. 7, a very high degree of correlation for the first few data points is observed. A window of 1000 data lag points is taken, which is around 8.33 h. Analyzing this plot for other components in other lines also showed a consistently high relation for up to 200 lags. This property can be leveraged by recurrent neural networks such as LSTM or GRU as they are excellent at learning long-term dependencies.

While multiple parameters have strong correlations with the harmonic component magnitudes, the goal is to design a system that takes the most basic inputs. These include time of the day and measured line currents for general models, and a self-sequence of fixed window for sequence-based models. This is done for the ease of implementing the system as complex measurements require costly equipment and introduce time delays into the system.

The data collected by the power analyzer is first extracted and arranged carefully in an organized *csv* file. This file is then imported as a dataset and analyzed thoroughly for patterns and correlations among the features. The inputs for different types of models are determined through the analysis. Now, these features are extracted into a separate dataset and grouped into training and testing subsets. The entire data workflow, from start to end, and the types of models used for comparison can be seen in Fig. 8. The outputs of the models are then concatenated together with actual values in a second $csv$ file which is fed into the Typhoon-HIL environment for visual validation using oscilloscopes.

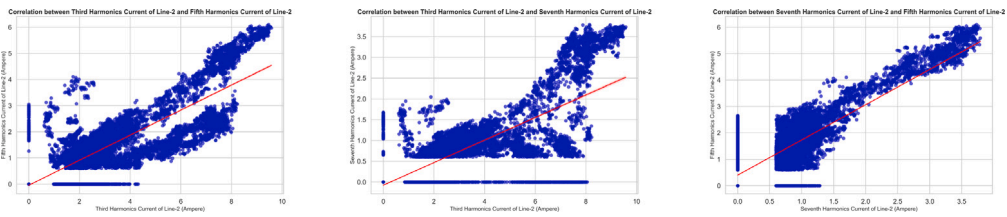

**Fig. 5.** Correlation plots of magnitudes of harmonic components with other harmonic component magnitudes of the same line (Ampere).

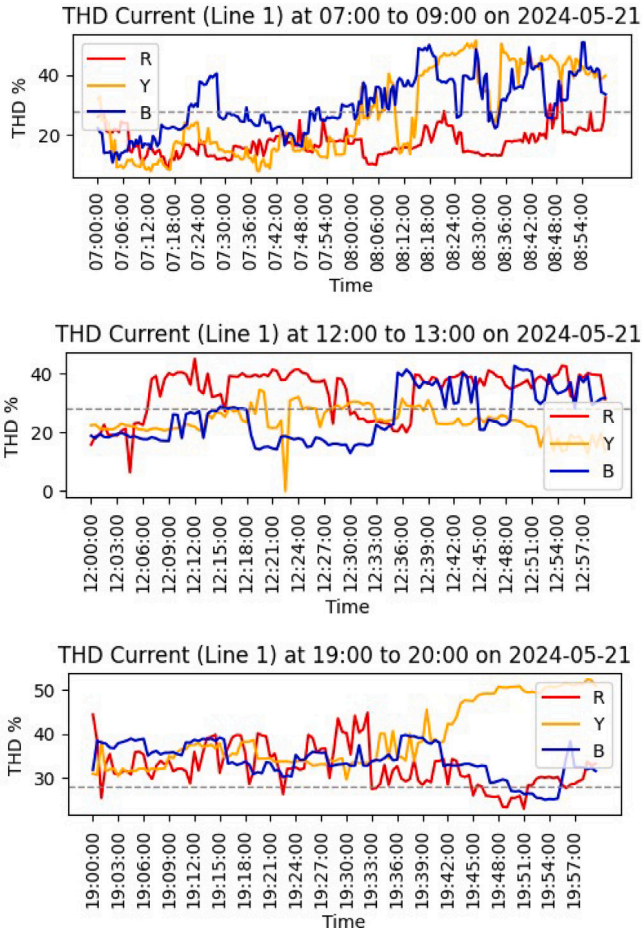

**Fig. 6.** Current THD for all three lines at different time windows for May 21.

### 3.3. Model architecture design

For this work, two distinct types of models are used: Ensemble models and Deep learning models. Ensemble models like $RandomForest$ and $GradientBooster$ perform well on a variety of datasets due to their nature of combining predictions from multiple base models. Within the Deep Learning architectures, basic Feed-forward Networks and Recurrent Networks are selected. The former is due to its ability to adapt to non-linear systems, while the latter is due to its capability to capture temporal dependencies.

Each model was initially designed to output three values: magnitudes of 3rd, 5th and 7th components. While it did simplify the training process, these models showed low accuracy. This may be due to the lack of a robust relationship among the harmonics, resulting in the trained parameters being optimized for one but not for the others, or simply subpar for all three. The predictions were found to be much more accurate when a single model was trained for a single component, keeping the overall architecture similar. This allows each model to learn the parameters for its particular component, instead of having to constantly adjust for all three at once. The models have been tuned using $KerasTuner$, a hyper-parameter optimization framework, along with additional manual adjustments. The description of each model architecture is followed by a table which includes the number and types of layers, the number of neurons in each layer, the connection of the layer and the total trainable parameters for each model. Each model is explained in the following sub-sections while all their parametric details are listed in Table 1.

#### 3.3.1. Ensemble methods

These are techniques that emphasize improving the accuracy and precision of the prediction results in machine learning models by combining multiple learning techniques. Under this learning methodology, a $RandomForest$ $Regressor$, available from $scikit-learn$ library and a $GradientBooster$ built from scratch are used.

Random forest is an ensemble method that uses bagging as the ensemble and a decision tree as the individual model. It is a flexible and easy-to-use algorithm which searches for the best feature, in a data set, from a random subset of provided features. On the other hand, Boosting is an ensemble technique that learns from previous predictor mistakes to enhance its future predictions. Gradient boosting adds a sequential component where preceding predictors correct their successors. This method utilizes gradient descent to identify and correct the errors. $RandomForest$ with 100 and 200 estimators are used respectively for Line 1 and Line 2 (all three harmonics through a single model), whereas, $GradientBooster$ with 100 estimators, made from scratch, is used for Line 3 with one model instance for each harmonic.

#### 3.3.2. Dense multi-layer perceptron

A Multi-Layer Perceptron is a type of feed-forward deep learning architecture where all the nodes in the previous and next layers are fully connected. The outputs of one layer are fed as the input to the next via a weight matrix multiplication. These weights are updated during the training session through backpropagation of error at the output node(s). An MLP is a simple yet powerful network in deep learning which can model non-linear systems easily via suitable activation functions. When using large networks, the chances of the model being over-fit are high. This refers to the situation when the model is overly adapted to the training set and cannot produce precise outputs for the testing/unknown sets. Regularization is a technique used to prevent overfitting by penalizing larger weights. Lambda ($\lambda$) is the regularizing parameter used in the first few layers, which has been set to 0.01 for this model.

#### 3.3.3. LSTM only

Long Short-Term Memory (LSTM) networks (Hochreiter and Schmidhuber, 1997) is a type of Recurrent Neural Network (RNN) capable of learning long-term dependencies. They are particularly well-suited for time-series forecasting and sequence prediction tasks. They use memory cells and gates to control the flow of information which allows them to selectively retain or discard previous information as needed, which addresses the vanishing gradient problem occurring in traditional RNNs. Nowadays, these are widely used in applications such as National Language Processing (NLPs) and time-series forecasting. The architecture of a single LSTM node, seen in Fig. 9, includes mathematical computation on three gated layers: forget gate, input gate and output gate. The forget gate, governed by Eq. (7), determines the information to be retained or discarded, from the previous hidden state ($h_{t-1}$) using a sigmoidal activation function. The input gate, given by Eqs. (8) and (9), adjusts the cell status using the current input and previous hidden state. This cell state is then output to the next node through Eq. (10). The overall LSTM node output is also transferred to the next node, given by Eq. (11) where $o_t$ is the output gate and $h_t$ is the overall LSTM node output.

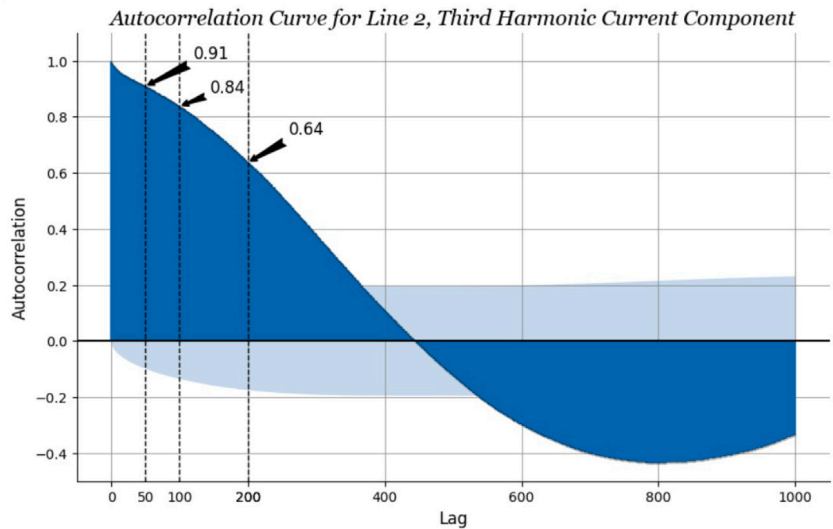

**Fig. 7.** Auto-correlation plot showing up to 1000 lags for the third harmonic magnitude.

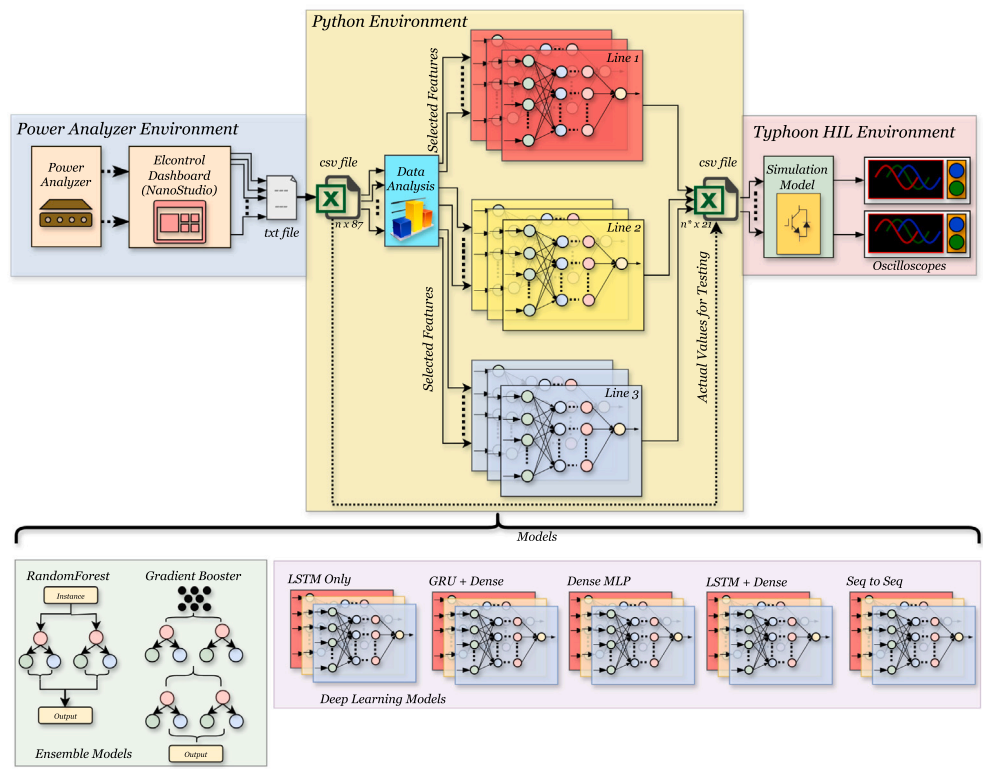

**Fig. 8.** Flow of data: starting at power analyzer environment for data collection, to python environment for analysis and model training to simulation environment for further observation in Typhoon HIL 404.

$$f_t = \sigma(W_f \cdot [h_{t-1}, x_t] + b_f) \tag{7}$$

$$i_t = \sigma(W_i \cdot [h_{t-1}, x_t] + b_i) \tag{8}$$

$$\tilde{c}_t = tanh(W_c \cdot [h_{t-1}, x_t] + b_c) \tag{9}$$

$$C_t = f_t \cdot c_{t-1} + i_t \cdot \tilde{c}_t \tag{10}$$

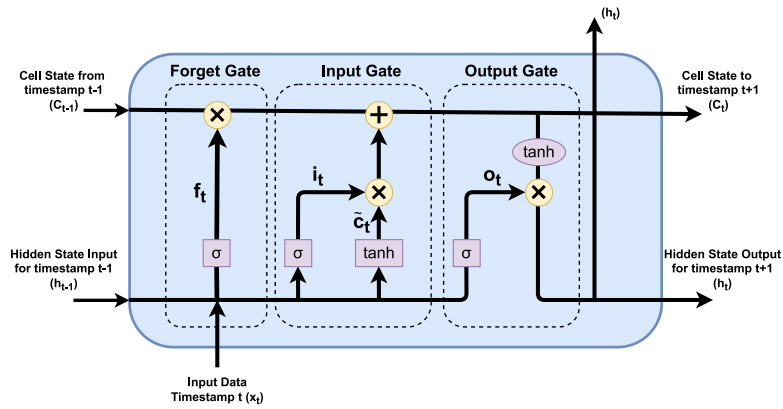

**Fig. 9.** Architecture of a single LSTM cell.

**Table 1**
Architecture and parameter details for all the models.

| Layer | Dense MLP | | LSTM Only | | LSTM + Dense | | GRU + Dense | | Seq-to-Seq | |
|---|---|---|---|---|---|---|---|---|---|---|
| | Type | Neurons | Type | Neurons | Type | Neurons | Type | Neurons | Type | Neurons |
| A | Dense | 1024 ($\lambda$) | LSTM | 352 | LSTM | 128 | GRU | 128 | LSTM | 64 |
| B | Dense | 512 ($\lambda$) | LSTM | 160 | Dropout | 128 | Dropout | 128 | LSTM | hco: 160 |
| C | Dense | 256 ($\lambda$) | LSTM | 128 | LSTM | 128 | GRU | 128 | R.V | 160 |
| D | Dense | 128 ($\lambda$) | Dropout | 128 | Flatten | 128 | Flatten | 128 | LSTM | 160 |
| E | Dense | 64 | – | – | Dense | 256 | Dense | 256 | Dropout | 160 |
| F | Dense | 32 | – | – | Dense | 128 | Dense | 128 | – | – |
| G | Dense | 8 | – | – | Dense | 64 | Dense | 64 | – | – |
| H | – | – | – | – | Dense | 32 | Dense | 32 | – | – |
| I | – | – | – | – | Dense | 16 | Dense | 16 | – | – |
| Output | Dense | 1 | Dense | 1 | Dense | 1 | Dense | 1 | Time-Dist | 1 |
| Trainable Params | 705 777 | | 974 849 | | 274 945 | | 226 177 | | 366 497 | |

$$o_t = \sigma(W_o \cdot [h_{t-1}, x_t] + b_o) \mid h_t = o_t \cdot tanh(C_t) \tag{11}$$

Since the harmonic components observed throughout the day are highly auto-correlative, LSTM can accurately predict the future sequences given the past. While the training time for an LSTM-based network is quite high, due to its complex node structure, it only requires a few epochs of training before giving precise outputs. This model requires a past sequence window input of 100 points.

### 3.3.4. LSTM + Dense

A network topology consisting of LSTM layers followed by Dense layers combines the strengths of both. LSTM layers are adept at recognizing and capturing dependencies and patterns in a given sequence. Using these recognized patterns, the fully connected (dense) layers map them to the required output space, enabling accurate and precise regression. This hybrid architecture offers an effective approach for sequence data such as the harmonic components within a measured time range.

### 3.3.5. GRU + Dense

Gated Recurrent Unit (GRU) (Cho et al., 2014) is a type of recurrent neural network with a simpler architecture compared to LSTM. The fusion of the input gate and forget gate into a single update gate leads to faster training while still being effective in capturing dependencies in the input. The Dense layers perform accurate regression to map the features to the outputs. The previously discussed LSTM + Dense architecture is utilized with the LSTM nodes replaced by GRU nodes. A GRU node has two gating mechanisms, called the reset gate ($r_t$) and update gate ($z_t$), as seen in Fig. 10. The reset gate, given in Eq. (12) determines how much of the previous state is forgotten. The update gate, given in Eq. (13) determines the effect of current input ($x_t$) while updating the hidden state. The candidate hidden state, given by Eq. (14) is then utilized alongside the update gate to determine the final output of the GRU node as per Eq. (15).

$$r_t = \sigma(W_r \cdot [h_{t-1}, x_t]) \tag{12}$$

$$z_t = \sigma(W_z \cdot [h_{t-1}, x_t]) \tag{13}$$

$$\tilde{h}_t = \tanh(W_h \cdot [r_t \cdot h_{t-1}, x_t]) \tag{14}$$

$$h_t = (1 - z_t) \cdot h_{t-1} + z_t \cdot \tilde{h}_t \tag{15}$$

### 3.3.6. Sequence to Sequence (Seq2seq)

This is a type of neural network design used when a sequence or pattern is provided as an input to predict future sequences. From auto-correlation curves, we can observe the cyclic pattern of harmonic components which can be considered as a sequence. In a typical seq2seq model, *encoders* process the input sequence and convert them into fixed-size vectors which is a summary of the sequence. *Decoders* take the summary vector and generate future output sequences. We observe a significant sequential correlation for all the harmonic components up to 200 units of lag. Considering architecture complexity, this model analyzes a past window of 100 lags to predict the next point in the sequence to reduce model complexity and training times. Fig. 11 shows the summarized architecture of the model used, in which there are a total of 150 hidden units. In Table 1, RV stands for Repeat Vector layer, which repeats the input a certain number of times (needed here for vector size matching) and Time-Dist refers to Time Distributed layer which applies a dense layer to each time-step slice of input.

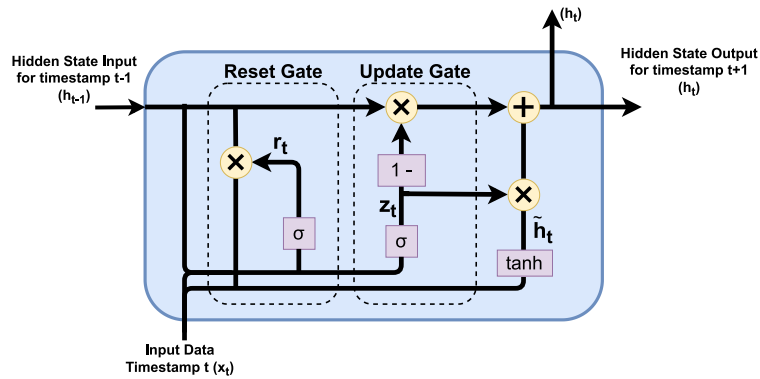

**Fig. 10.** Architecture of a single GRU cell.

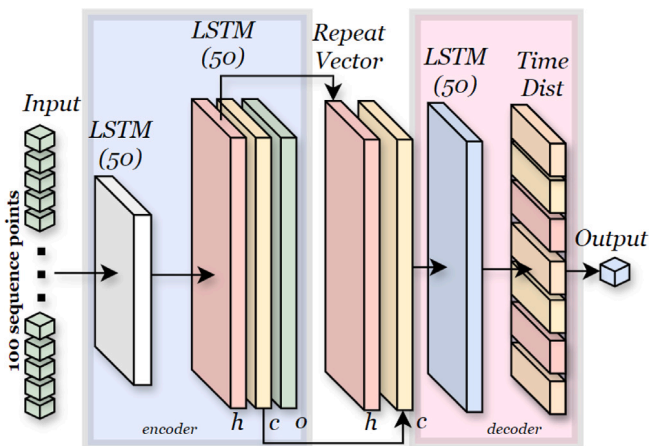

**Fig. 11.** A typical sequence-to-sequence model architecture representation.

### 3.4. Training

The dataset is divided into three subsets: training set, test set and validation set. Each of the architectures is trained separately using the training set, with different hyper-parameters. These are determined through experimentation as well as a hypertuner called *keras-tuner*. It is a scalable hyperparameter optimization framework that tries to improve the performance of a model by adjusting certain values, such as the number of nodes in a layer. The training times depend on the complexities of each architecture. Simple MLPs with only fully connected layers are much faster to train compared to sequence-to-sequence networks, however, they require more iterations. Additionally, it is not always given that the best model is obtained at the end of the training process. To account for this, model checkpoints are used to ensure models with the best-fitting weights are saved instantly. It does so by monitoring the performance after each iteration of the validation set. Once this process is over, the saved files are called (used) to produce predictions for unseen inputs in the test set.

### 3.5. Simulation

A simplified simulation is carried out in Typhoon HIL to visually verify the improvement of waveforms due to harmonic cancellation using predicted component magnitudes. Typhoon HIL is a real-time simulation platform specifically designed for hardware-in-the-loop (HIL) testing, enabling the accurate modeling and validation of power systems in real-time. It allows dynamic testing by simulating physical environments without the need for actual hardware prototypes, ensuring safer and faster development. The simulation is first set up and then loaded into the HIL device. A separate SCADA environment is set up to observe the key parameters in real-time while the simulation runs.

In the simulation setup, the Typhoon HIL 404 is directly connected to a laptop, which serves as the host machine to develop the system model. These models are created using specialized Typhoon HIL software on the laptop, that allows the designing, testing, and simulation of the system virtually. Once the system model is complete and validated within the simulation environment, it is loaded onto the HIL device for real-time testing. The breakout board, which is a hardware interface that provides access to the input/output (I/O) ports of the HIL device, is connected to the back of the Typhoon HIL 404. This enables interaction between the simulator and other external components. The output ports on the breakout board are configured to correspond to specific signals generated by the simulation. These signals are mapped to specific pins on the board, which are then connected to external monitoring devices like an oscilloscope. The mapped outputs, such as voltage waveforms or Pulse Width Modulation (PWM) signals are the key parameters and can be observed in real-time both on the oscilloscope for physical signal verification and within the Typhoon HIL-SCADA system.

The overall filter, as seen in Fig. 12, is a simple hysteresis band-controlled inverter with a capacitor connected on its DC side. The DC side voltage is used as a reference to adjust the signal magnitudes. Hysteresis band is a simple yet sturdy control method to generate Pulse Width Modulation (PWM) signals based on generated reference signals. It consists of relays that ensure the instantaneous output values oscillate within a fixed band around the reference signal. The measured and predicted harmonic magnitudes are utilized as inputs for simulation. A macro function, defined within the SCADA before running the simulation, allows the reading of the CSV file where the data is stored. Our survey showed a substantially higher THD in current waveforms. Signal-controlled current sources are used to imitate the exact conditions, using the actual values of fundamental and harmonic components. Reference signals are generated by the sum of three harmonic components, for each of the phases. The magnitudes of these components are the predicted values of the model architectures. This is then inverted and passed through the hysteresis band controller which generates the PWM signals. These signals control the switching of the inverter legs.

The purpose of the simulation is to validate the waveform filter visually through the oscilloscope. Its implementation in an actual power system will require further adjustments.

## 4. Results

After the training process, each model is evaluated on the test dataset. This set acts as an "unseen" input for the model, for which it generates predictions. A test dataset is crucial for evaluation as it provides a way to evaluate the models using unique cases with the ability to compare predictions with target output values.

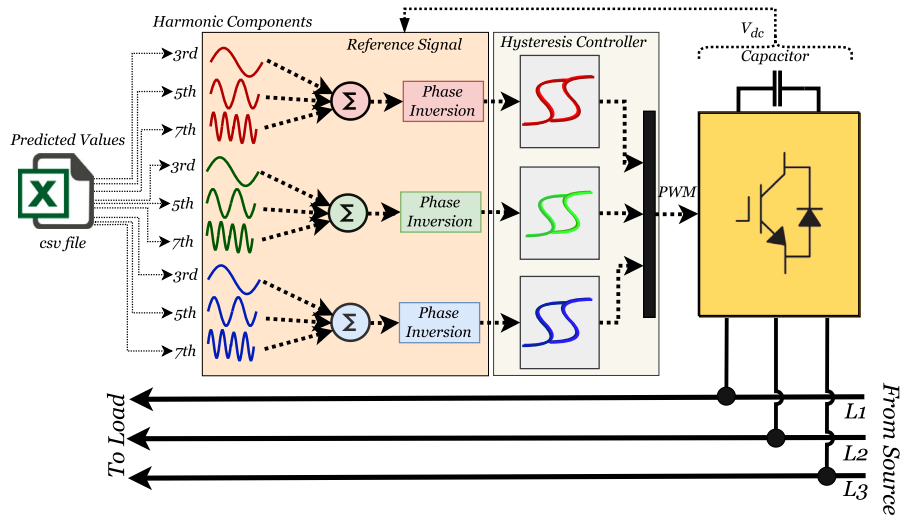

**Fig. 12.** Basic control of the filter inverter using predicted harmonic components, waveform generation and hysteresis band control.

**Fig. 13.** Features extracted in a csv file with fundamental, actual (Act) and predicted (Pred) harmonics magnitudes for each line.

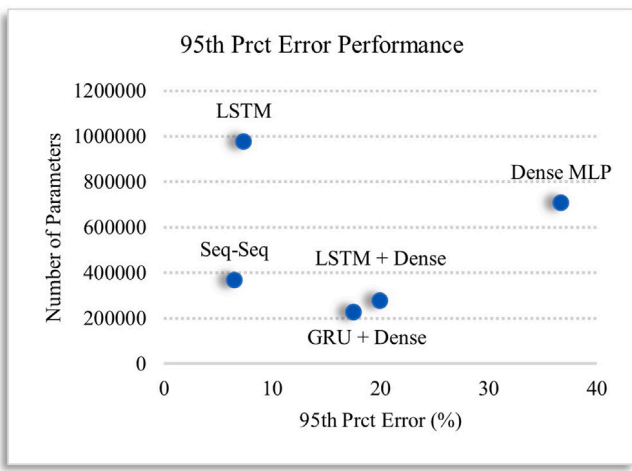

**Fig. 14.** 95th Percentile error for each model on the testing dataset.

The performance of any regressive neural network is determined by how accurate its predictions are. In our case, a model that can predict the magnitudes of harmonic components with the least deviation from the actual value is the one that performs the best. Relative error, given by Eq. (16), is used to test the models. The lower the error, the better the performance.

The predicted values are then concatenated with the actual ones to form a feature list as seen in Fig. 13. Here, $Fnd$ represents the fundamental current magnitude for a particular line while $Act$ and $Pred$ represent actual and predicted component values for each line. All the values are in Amperes since they all represent currents. This arrangement is the input file for the HIL simulation.

$$Relative\ Error(\%) = \frac{|actual - predicted|}{actual} * 100\% \qquad (16)$$

Fig. 14 shows the comparison of the deep learning models: based on 95th percentile error and model size. Using the predictions outputted by the best-performing model (seq2seq), harmonic cancellation can be observed in the oscilloscopes as seen in Fig. 15. The figure shows two random cases where THDi is considerably high (>20%) which is reduced to under 10% through harmonic cancellation. The harmonic distortion, in the simulation, is measured using a THD measurement block. The inherent controlling mechanism of a hysteresis band controller injects some harmonics, which may reflect on the THDi values measured post-filter.

Fig. 16 shows the pre and post-filter parameters for 10 different cases in Typhoon HIL-SCADA simulation environment. These values are then tabulated in Table 2. For each case, pre-filter parameters are on the left while the post-filter parameters are shown on the right. On the sides of the waveforms, THDi (first value on each side), and the magnitudes in Ampere for the actual (left) and predicted (right) harmonic components (3rd, 5th and 7th sequentially) can be seen. We can observe proper filtering for most cases, while some cases show little improvements. Tt was observed that filtering was ineffective mostly when the line was lightly loaded. This could be because, under such cases, the effects of even slight errors in magnitudes are magnified.

A histogram plot can be used to visualize the loss of each model. The plot, as seen in Fig. 17, is labeled to show the average error for each instance. Loss implies the difference between the actual target output and the predicted output by the model. Lower losses signify high accuracy of prediction and are a crucial factor in comparing different architectures. It is clear from the histograms and the 95th percentile error that the Seq2Seq model performs the best while also having a high number of parameters. This can be credited to the high auto-correlation observed for each harmonic component. This architecture can maintain and propagate hidden states through time, leveraging LSTM nodes that can effectively manage long-term dependencies. Such networks use gating mechanisms to selectively retain relevant information from previous time steps, which allows them to capture the strong temporal correlations present in auto-correlated data. Additionally, an encoder–decoder structure enhances its ability to focus on important parts of the sequence, improving predictive accuracy. The predictions made by the Seq2Seq model were tabulated and used for the simulation.

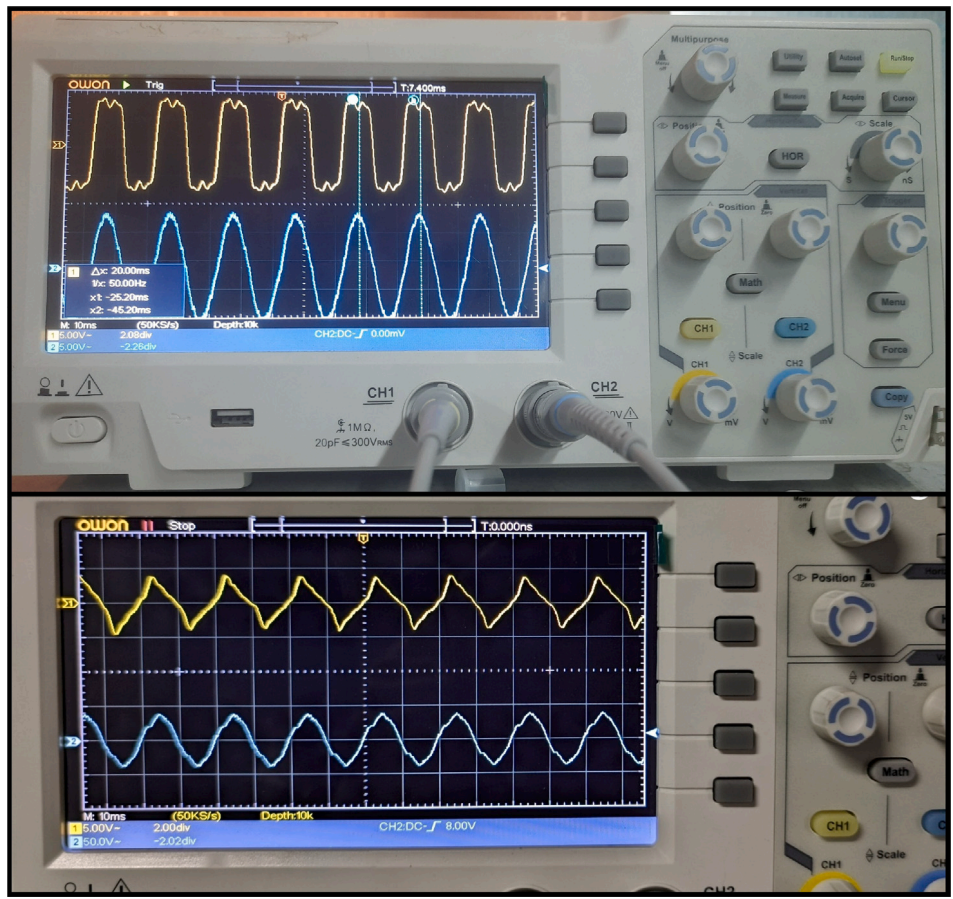

**Fig. 15.** Waveform before (yellow) and after (blue) harmonic cancellation for two randomly selected cases.

**Table 2**
Observed parameters for the ten different cases presented in Fig. 16.

| Case | Third harmonic (Ampere) | | Fifth harmonic (Ampere) | | Seventh harmonic (Ampere) | | THDi (%) | |
|------|--------|-----------|--------|-----------|--------|-----------|------------|-------------|
| | Actual | Predicted | Actual | Predicted | Actual | Predicted | Pre filter | Post Filter |
| Case 1 | 3.71 | 3.78 | 1.39 | 1.37 | 0.69 | 0.68 | 15% | 5% |
| Case 2 | 3.78 | 3.75 | 0.73 | 1.33 | 0.73 | 0.87 | 31% | 9% |
| Case 3 | 5.08 | 4.84 | 2.16 | 2.22 | 0.00 | 0.00 | 33% | 5% |
| Case 4 | 9.31 | 9.49 | 2.57 | 2.39 | 0.00 | 0.01 | 48% | 3% |
| Case 5 | 6.67 | 5.07 | 1.10 | 1.14 | 0.79 | 0.77 | 44% | 3% |
| Case 6 | 6.51 | 6.61 | 0.00 | 0.04 | 0.80 | 0.94 | 35% | 7% |
| Case 7 | 5.06 | 4.86 | 2.80 | 2.63 | 1.60 | 1.59 | 28% | 7% |
| Case 8 | 5.42 | 5.31 | 1.32 | 1.45 | 0.74 | 0.73 | 26% | 5% |
| Case 9 | 3.93 | 3.74 | 1.32 | 1.23 | 0.70 | 0.71 | 18% | 6% |
| Case 10 | 2.84 | 2.85 | 4.97 | 4.92 | 1.13 | 1.21 | 20% | 14% |

The ensemble models have also shown great accuracies. They perform well due to their nature of combining predictions from multiple base models, reducing the risk of over-fitting and increasing generalization. By aggregating the strengths of diverse models, ensembles capture different patterns and decision boundaries that individual models might miss.

Similar work relating to Harmonic prediction and suppression has been done in Iqbal et al. (2021). The authors have designed a neural network-based Shunt Hybrid Active Power Filter (SHAPF) to enhance power quality. The best results have been obtained using LSTM-based RNN. The authors have been able to reduce the THDi of the system from as high as 37% and above to around 4%. In a few exceptional cases, for simple ANN networks, the THDi has improved by only a small margin of less than 5%. Similar to our result, the authors have concluded that recurrent networks that are capable of capturing temporal dependencies performed the best.

## 5. Discussion and conclusion

With the increase in the use of power electronic devices, there is a growing emphasis on improving power quality. The traditional methods, while robust, are quite slow and require complex calculations. The recent boom in Artificial Intelligence has highlighted the capability of neural networks in modeling nonlinear systems to produce accurate outputs. This proposed method explores various neural network architectures, leveraging their characteristics to predict harmonic components in any distribution line. The instantaneous and accurate predictions fundamentally remove any concerns of errors and time delay. The periodicity of the harmonic components, observed during data analysis is further validated by the strong performance of the sequential architecture.

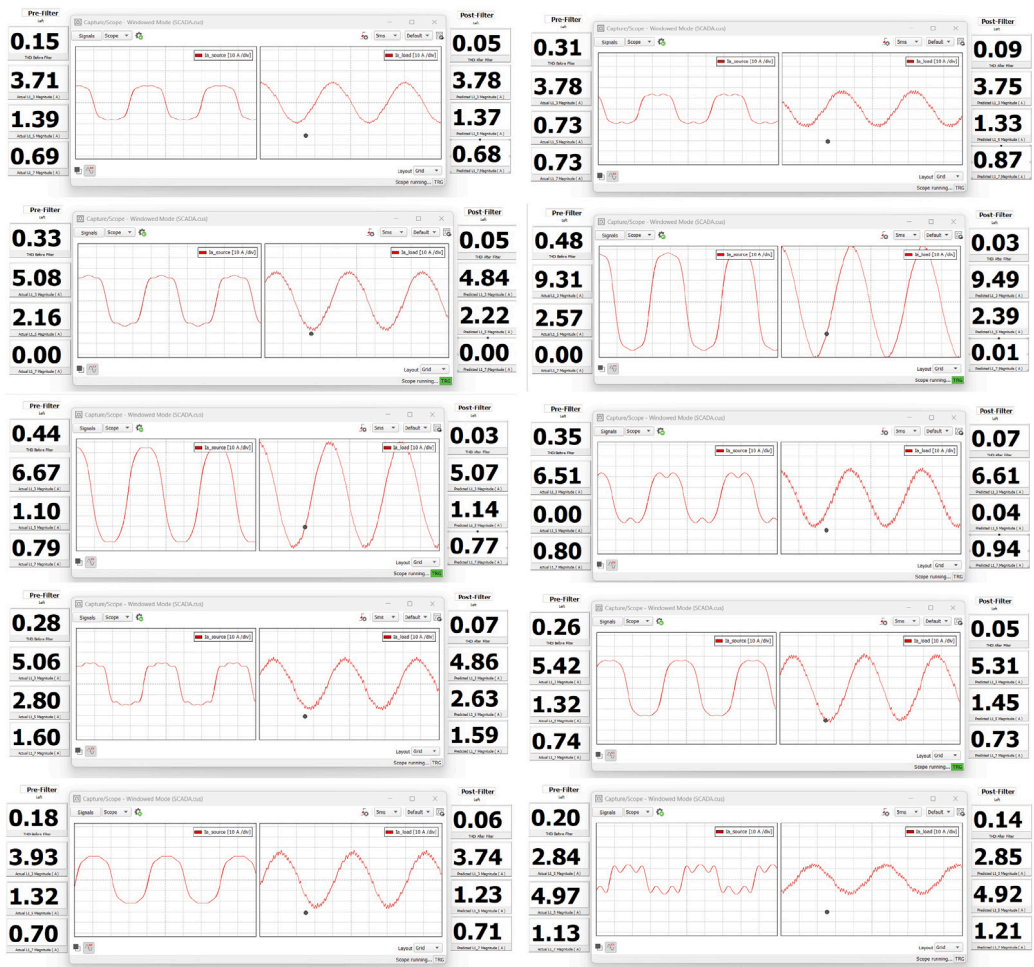

**Fig. 16.** Ten random test cases in HIL SCADA showing Line 1 parameters: before filter waveform, THDi and harmonic magnitudes (left) and after filter waveform, THDi and harmonic magnitudes (right).

This methodology, once adapted practically, can be used for harmonic component prediction for filters in small-scale industrial areas or residential areas. Since the architectures have to be trained with relevant data first, it is easier in smaller locations. These locations tend to have a pattern in their load consumption. Using transfer learning and generalized networks, such models can then be adapted for larger distribution systems, that may exhibit a more complex and wayward nature.

With data being at the center of the modern generation, neural network-based control systems will likely replace the traditional ones due to their high accuracy, adaptability and instantaneous response. The capability of an ANN-based approach to predict future harmonic components to aid in filtering is quite effective. Analyzing the trend and pattern in the harmonic components, while also using easily measurable parameters such as current and time as inputs, highly accurate model architectures were built. The results some models showed were highly promising, with THDi reductions of up to 30%. The distribution system of developing nations like Nepal has high harmonic components, due to the extensive use of power electronic devices with low-quality filters. Leveraging instantaneous predictions, active filters can be implemented at various distribution points where measurements can be readily taken. Adding constraints on the model outputs can help the model deal with anomalies, which can be implemented through self-defined unique activation functions.

The availability of relevant and accurate data is the biggest concern for our proposed methodology. If not trained properly, the network could be susceptible to extreme variability in real-life power systems caused by unexpected load changes, power outages and system instability. Additionally, for a continuous control, the datasets would be enormous. Also, as we transition to a digital and data-heavy power system, the potential for hacking and data theft is imminent. Addressing these risks will be crucial in the implementation of data-based control systems in general.

Improvements for this approach include implementation of the filter for a voltage source network and analysis of the system viability with physics-informed neural networks and other advanced neural-network architectures. Real-time training, once implemented practically, will allow the model to constantly adapt its parameters ensuring constant improvement of its predictions with time. Furthermore, coupling this system with ANN-based topologies for other elements, such as PLLs, could result in an even more robust control. The introduction of load/demand parameters of a particular area where the system is to be connected could be beneficial in reinforcing the system for long-term usage. The usage of neural-network predicted data in the stable operation of physical systems must be studied and optimized through constraints and parameter boundaries. Real-time data manipulation and control will highly benefit from technological advancements in processing power.

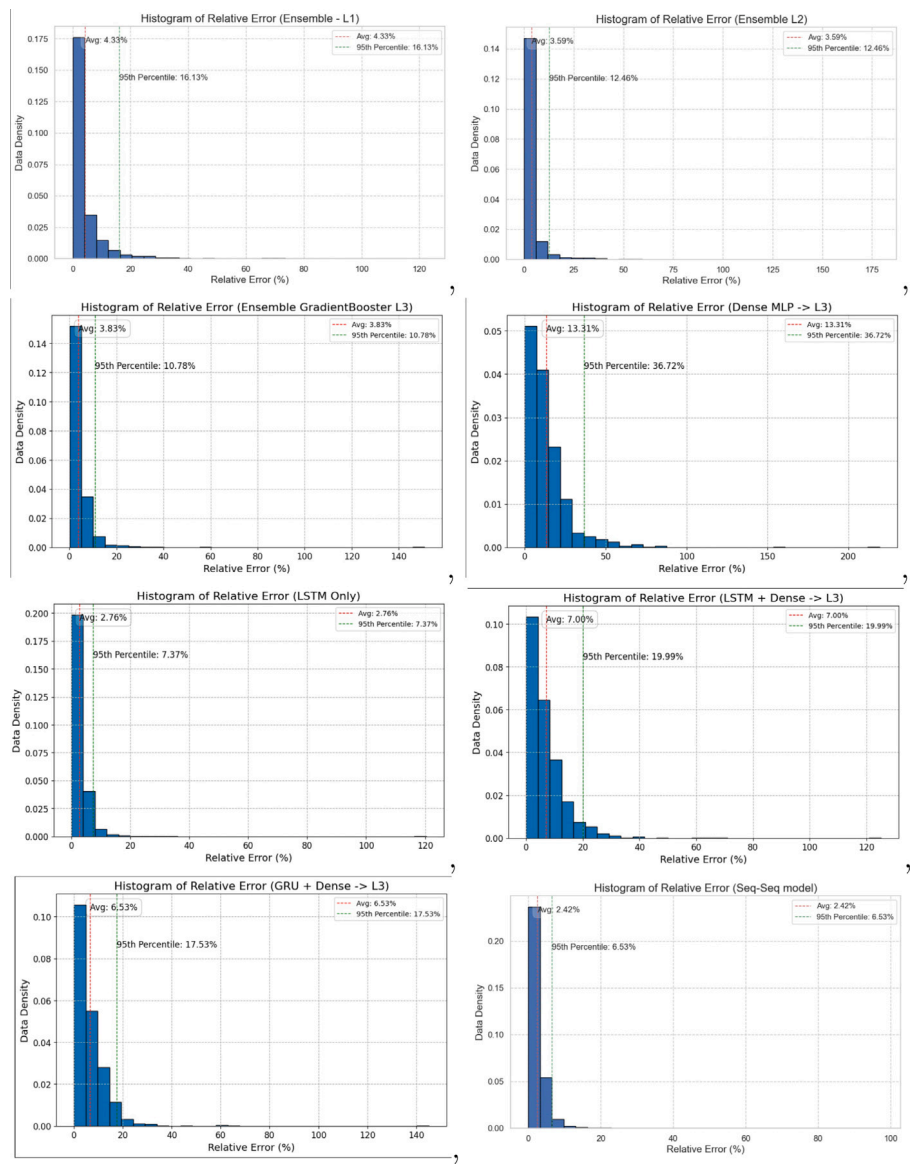

**Fig. 17.** Error Histogram depicting the testing results for each of the models, showing the average and 95th percentile error.

## CRediT authorship contribution statement

**Dixant Bikal Sapkota:** Writing – original draft, Visualization, Methodology, Data curation, Conceptualization. **Puskar Neupane:** Writing – original draft, Visualization, Methodology, Formal analysis, Data curation. **Kajal Pokharel:** Writing – review & editing, Visualization, Data curation. **Shahabuddin Khan:** Writing – review & editing, Validation, Supervision, Resources, Project administration, Funding acquisition.

## Declaration of competing interest

The authors declare that they have no known competing financial interests or personal relationships that could have appeared to influence the work reported in this paper.

## Data availability

Data will be made available on request.

## Acknowledgement

This work was supported by the Nepal Academy of Science and Technology (NAST).

## References

Afonso, J.L., Tanta, M., Pinto, J.G.O., Monteiro, L.F.C., Machado, L., Sousa, T.J.C., Monteiro, V., 2021. A review on power electronics technologies for power quality improvement. Energies 14 (24).

Chaudhari et al, M., 2012. Three-phase series active power filter as power quality conditioner. In: 2012 IEEE International Conference on Power Electronics, Drives and Energy Systems. PEDES, IEEE, pp. 1–6.

Arrillaga, J., Watson, N., 2003. Power System Harmonics. Wiley InterScience electronic collection, Wiley.

Bermejo, J., Gomez Fernandez, J.F., Polo, F., Crespo Marquez, A., 1844. A review of the use of artificial neural network models for energy and reliability prediction. a study of the solar pv, hydraulic and wind energy sources. Appl. Sci. 9, 052019.

Chen, S., Billings, S.A., 1992. Neural networks for nonlinear dynamic system modelling and identification. Internat. J. Control 56 (2), 319–346.

Chen, X., Qu, G., Tang, Y., Low, S., Li, N., 2022. Reinforcement learning for selective key applications in power systems: Recent advances and future challenges. IEEE Trans. Smart Grid 13, 2935–2958.

Cho, K., van Merrienboer, B., Bahdanau, D., Bengio, Y., 2014. On the properties of neural machine translation: Encoder–decoder approaches.

Costa-Castelló, R., Griñó, R., Parpal, R.C., Fossas, E., 2009. High-performance control of a single-phase shunt active filter. IEEE Trans. Control Syst. Technol. 17 (6), 1318–1329.

Dixon, J.W., Venegas, G., Moran, L.A., 1997. A series active power filter based on a sinusoidal current-controlled voltage-source inverter. IEEE Trans. Ind. Electron. 44 (5), 612–620.

do Nascimento, C.F., de Oliveira, A.A., Goedtel, A., Amaral Serni, P.J., 2011. Harmonic identification using parallel neural networks in single-phase systems. Appl. Soft Comput. 11 (2), 2178–2185, The Impact of Soft Computing for the Progress of Artificial Intelligence.

Dugan, R., Santoso, S., McGranaghan, M., Beaty, H., 2002. Electrical Power Systems Quality. McGraw-Hill Professional Engineering, McGraw-Hill Education.

El-Mamlouk, W.M., Mostafa, H.E., El-Sharkawy, M.A., 2011. Active power filter controller for harmonic suppression in industrial distribution system. Ain Shams Eng. J. 2 (3), 161–172.

Fabricio, E.L.L., Júnior, S.C.S., Jacobina, C.B., de Rossiter Corrêa, M.B., 2018. Analysis of main topologies of shunt active power filters applied to four-wire systems. IEEE Trans. Power Electron. 33 (3), 2100–2112.

Fuchs, E.F., Masoum, M.A., 2023. Chapter 10 - the roles of filters in power systems and unified power quality conditioners. In: Fuchs, E.F., Masoum, M.A. (Eds.), Power Quality in Power Systems, Electrical Machines, and Power-Electronic Drives (Third Edition), third ed. Academic Press, pp. 915–1016.

George, S., Agarwal, V., 2006. A dsp-based control algorithm for series active filter for optimized compensation under nonsinusoidal and unbalanced voltage conditions. IEEE Trans. Power Deliv. 22 (1), 302–310.

Guzman, R., de Vicuña, L.G., Morales, J., Castilla, M., Miret, J., 2016. Model-based control for a three-phase shunt active power filter. IEEE Trans. Ind. Electron. 63 (7), 3998–4007.

Hochreiter, S., Schmidhuber, J., 1997. Long short-term memory. Neural Comput. 9 (8), 1735–1780.

Iqbal, M., Jawad, M., Jaffery, M.H., Akhtar, S., Rafiq, M.N., Qureshi, M.B., Ansari, A.R., Nawaz, R., 2021. Neural networks based shunt hybrid active power filter for harmonic elimination. IEEE Access 9, 69913–69925.

Javadi, A., Hamadi, A., Woodward, L., Al-Haddad, K., 2016. Experimental investigation on a hybrid series active power compensator to improve power quality of typical households. IEEE Trans. Ind. Electron. 63 (8), 4849–4859.

Jiao, S., Potti, K.R.R., Rajashekara, K., Pramanick, S.K., 2019. A novel drogi-based detection scheme for power quality improvement using four-leg converter under unbalanced loads. IEEE Trans. Ind. Appl. 56 (1), 815–825.

Jou, H.-L., Wu, J.-C., Chu, H.-Y., 1994. New single-phase active power filter. IEE Proc., Electr. Power Appl. 141 (3), 129–134.

Kanjiya, P., Khadkikar, V., Zeineldin, H.H., 2014. Optimal control of shunt active power filter to meet ieee std. 519 current harmonic constraints under nonideal supply condition. IEEE Trans. Ind. Electron. 62 (2), 724–734.

Kim, S., Kim, J., Ko, S., 2004. Three-phase three-wire series active power filter, which compensates for harmonics and reactive power. IEE Proc., Electr. Power Appl. 151 (3), 276–282.

Lightowler, N., Nareid, H., 2003. Artificial neural network based control systems. SAE Trans. 112, 539–543.

Lin, H.C., 2007. Intelligent neural network-based fast power system harmonic detection. IEEE Trans. Ind. Electron. 54 (1), 43–52.

Mohan, N., Undeland, T., Robbins, W., 2003. Power Electronics: Converters, Applications, and Design. No. V. 1 in Power Electronics: Converters, Applications, and Design. John Wiley & Sons.

Nejdawi, I., Emanuel, A., Pileggi, D., Corridori, M., Archambeault, R., 1999. Harmonics trend in ne usa: a preliminary survey. IEEE Trans. Power Deliv. 14 (4), 1488–1494.

Olivencia Polo, F.A., Ferrero Bermejo, J., Gómez Fernández, J.F., Crespo Márquez, A., 2015. Failure mode prediction and energy forecasting of pv plants to assist dynamic maintenance tasks by ann based models. Renew. Energy 81, 227–238.

Pandey, U., Pathak, A., Kumar, A., Mondal, S., 2023. Applications of artificial intelligence in power system operation, control and planning: a review. Clean Energy 7, 1199–1218. http://dx.doi.org/10.1093/ce/zkad061.

Pinto, J.G., Carneiro, H., Exposto, B., Couto, C., Afonso, J.L., 2011. Transformerless series active power filter to compensate voltage disturbances. In: Proceedings of the 2011 14th European Conference on Power Electronics and Applications. pp. 1–6.

Qasim, M., Kanjiya, P., Khadkikar, V., 2014. Artificial-neural-network-based phase-locking scheme for active power filters. IEEE Trans. Ind. Electron. 61 (8), 3857–3866.

Rice, D.E., 1986. Adjustable speed drive and power rectifier harmonics-their effect on power systems components. IEEE Trans. Ind. Appl. IA-22 (1), 161–177.

Rosenblatt, F., 1958. The perceptron: a probabilistic model for information storage and organization in the brain. Psychol. Rev. 65 (6), 386.

Sapkota, D.B., Neupane, P., Joshi, M., Khan, S., 2024. Deep learning model for enhanced power loss prediction in the frequency domain for magnetic materials, IET Power Electronics.

Sasaki, H., Machida, T., 1971. A new method to eliminate ac harmonic currents by magnetic flux compensation-considerations on basic design. IEEE Trans. Power Appar. Syst. (5), 2009–2019.

Schöning, J., Riechmann, A., Pfisterer, H.-J., 2022. Ai for closed-loop control systems: New opportunities for modeling, designing, and tuning control systems. In: Proceedings of the 2022 14th International Conference on Machine Learning and Computing. ICMLC '22, Association for Computing Machinery, New York, NY, USA, pp. 318–323.

Shatshat, R., Salama, M., Kazerani, M., 2004. Artificial intelligent controller for current source converter-based modular active power filters. IEEE Trans. Power Deliv. 19 (3), 1314–1320.

Suykens, J.A., Vandewalle, J.P., De Moor, B.L., 2012. Artificial Neural Networks for Modelling and Control of Non-Linear Systems. Springer Science & Business Media.

Taghvaie, A., Warnakulasuriya, T., Kumar, D., Zare, F., Sharma, R., Vilathgamuwa, D., 2023. A comprehensive review of harmonic issues and estimation techniques in power system networks based on traditional and artificial intelligence/machine learning. IEEE Access 11, 31417–31442.

Verdelho, P., Marques, G.D., 1998. Four-wire current-regulated pwm voltage converter. IEEE Trans. Ind. Electron. 45 (5), 761–770.

Wang, Y., 2024. Research on harmonic suppression methods. Highlights Sci. Eng. Technol. 87, 138–142.

Waterworth, G., Lees, M., 2000. Artificial neural networks in the modelling and control of non-linear systems. In: IFAC Proceedings Volumes. Vol. 33, pp. 95–97, 1. 2000. IFAC Workshop on Programmable Devices and Systems. PDS 2000, Czech Republic, Ostrava, p. 8-9.

Zhang, G., Eddy Patuwo, B., Hu, M.Y., 1998. Forecasting with artificial neural networks:: The state of the art. Int. J. Forecast. 14 (1), 35–62.